\documentclass[aps,twocolumn,showpacs,preprintnumbers]{revtex4-2}
\usepackage{float}
\usepackage{dblfloatfix}
\usepackage{fixltx2e}
\restylefloat{table}
\usepackage{graphicx}  
\usepackage{subfig}
\usepackage{subfloat}
\usepackage{multirow}
\usepackage{fancyhdr}
\usepackage{parskip}
\usepackage[T1]{fontenc}
\usepackage{dcolumn}   
\usepackage{bm}        
\usepackage{amsfonts}  
\usepackage{amsmath}   
\usepackage{amssymb}   
\usepackage{makecell}
\usepackage{caption}
\usepackage{xcolor}

\begin{document}

\title{Thermal equilibrium of a Brownian particle with coordinate dependent diffusion: comparison of the Boltzmann and a modified Boltzmann distribution with experimental results}

\author{Abhinav Dhawan}
\author{A. Bhattacharyay}
\email{a.bhattacharyay@iiserpune.ac.in}

\affiliation {\it Indian Institute of Science Education and Research, Pune, India}

\date{June 7, 2022}

\begin{abstract}  
In this paper we compare the Boltzmann distribution with a modified Boltzmann distribution, that results from an It\^o-process considering thermal equilibrium of a Brownian particle with coordinate dependent diffusion, in the light of an existing experiment. The experiment was reported in 1994 by Faucheux and Libchaber. The experiment made use of direct tracking of diffusion of Brownian particles near a wall. Results of this experiment allows us to compare the Boltzmann and the modified Boltzmann distribution without making use of any adjustable parameter. A comparison of these two distributions with the experimental results lends support to the consideration of thermodynamic equilibrium of a Brownian particle with coordinate-dependent diffusion to be an It\^o-process.

\end{abstract}

\pacs{05.10.Gg, 05.20.-y,05.40.Jc,05.70.-a}

\maketitle 

\section{Introduction}
Understanding position distribution of a confined Brownian particle (BP) with coordinate dependent diffusion and in thermal equilibrium with a heat-bath of constant temperature is important due to immense possibility of its application \cite{roussel2004reaction,barik2005quantum,sargsyan2007coordinate,chahine2007configuration,best2010coordinate,lai2014exploring,berezhkovskii2017communication,foster2018probing,ghysels2017position, yamilov2014position} across disciplines. A BP near a wall or, similarly, near other BP's see coordinate dependence of effective-diffusion as a rule than an exception. The underlying reason of this coordinate (or space) dependence of diffusion is considered mostly to be hydrodynamic. The phrase effective-diffusion, in the present context, indicates local renormalization of the diffusivity of a BP by the average presence of hydrodynamic modes due to the proximity to walls and interfaces.

When it is possible to define/measure the coordinate dependence (and time independence) of diffusivity of a BP, the consideration is that the BP, at the least locally, has thermally equilibrated with the heat-bath. This spatially inhomogeneous Brownian motion models a slower local isotropic transport of a much larger BP in comparison to the time and length and scales of the bath degrees of freedom. These bath degrees of freedom could be those faster hydrodynamic modes whose average effect is modeled by Brownian motion. This is the standard wisdom which even underlies the existence of coordinate-dependent diffusivity because the local method of measurement is the same using mean-square-displacement (MSD) against time. Due to this definition/measurement of diffusivity the transport it represents is always locally isotropic even when the diffusivity is coordinate-dependent.  

It is important to note that, only the locally isotropic part of the average local-transport caused by hydrodynamic effects should be modeled by a diffusion process which has space-dependent diffusivity. The locally anisotropic part of the average transport needs the recognition of an effective force to account for the drift it would produce in the context of the Langevin dynamics. If one tries to model existence of such a drift due to hydrodynamic reasons by a noise term in the Langevin dynamics, one would end up using correlated noise and the thermal nature of the noise will get distorted. This anisotropic part of local transport due to hydrodynamic effects is of no concern to us in the present context and we assume here that the process does not involve such a local drift unaccounted for by the known interactions. It is absolutely technically possible to keep the diffusive transport isotropic at every point in space even when the diffusivity changes from point to point and that is the key consideration underlying the presence of an uncorrelated thermal noise modelling Brownian motion.

Given this context, the question that may arise is - what is the spatial distribution of such a BP over an inhomogeneous space where the source of inhomogeneity (space-dependent diffusivity) cannot feature in the Hamiltonian of the system? This question has not been paid much attention to until recently. Bhattacharyay and coworkers \cite{bhattacharyay2019equilibrium,bhattacharyay2020generalization,maniar2021random} have shown that, for thermodynamic equilibrium, such a coordinate-dependent diffusion process is essentially an It\^o-process and that gives a modified Boltzmann distribution (MBD) for thermal equilibrium. The resulting MBD of an It\^o-process can be written (for example in one-dimensional space) as
\begin{equation}\nonumber
    P(x) = \frac{N}{D(x)}e^{-U(x)/[D(x)\Gamma(x)]},
\end{equation}
where $D(x)$ is the coordinate-dependent diffusivity and $\Gamma(x)$ is the coordinate-dependent damping coefficient. The interaction potential that confines the particle to equilibrate with a heat-bath is $U(x)$. Obviously, in the above expression, $P(x)$ is the probability density at coordinate $x$ where $N$ is the normalization constant. The Langevin dynamics does not in any way ensure local validity of the Stokes-Einstein relation $D(x)\Gamma(x)=k_BT$ where $k_B$ is the Boltzmann constant and $T$ is the temperature. The Stokes-Einstein relation is always imposed to the Langevin dynamics. If one does so to an It\^o-process, one would get the Boltzmann factor appear in the expression of the MBD. It has been shown that the Fick's law as it comes from the Kramers-Moyal expansion is the correct one to consider \cite{bhattacharyay2019equilibrium} in any convention (It\^o or Stratonovich) and the Stokes-Einstein relation corresponding to a constant temperature everywhere in space is consistent with an It\^o process \cite{bhattacharyay2020generalization}. 

Bhattacharyay has shown that the equilibrium distribution of a generalized Langevin dynamics (involving inertial term and velocity-distribution thereof) under the local validity of Stokes-Einstein relation has the over-damped limit which converges to the equilibrium distribution of an It\^o-process \cite{bhattacharyay2020generalization}. This is a crucial observation in this context because an over-damped dynamics (corresponding, for example, to an It\^o-process) does not contain any information of velocity distribution and, hence, temperature. Temperature is imposed to an over-damped dynamics by the imposition of the Stokes-Einstein relation. One has to, therefore, understand if at all the imposition of the Stokes-Einstein relation is consistent with the stochastic process in its over-damped limit. These observations have been further extended to random-walk methods developed by Maniar and Bhattacharyay in \cite{maniar2021random} where the maintenance of a constant temperature everywhere in space by the local-validity of the Stokes-Einstein relation is confirmed. Such a system is seen to correspond to an It\^o-process at the continuum limit of the random walk \cite{maniar2021random}.

On the contrary, theoretical imposition of the Boltzmann distribution (BD) has been the basis of theoretically considering equilibrium in the context of coordinate-dependent diffusion in existing literature so far \cite{sokolov2010ito,lau2007state,sancho1982adiabatic,sancho2011brownian,farago2014langevin,farago2014fluctuation}. In this connection, to get the Boltzmann distribution for such a BP, one treats the multiplicative noise problem using the Stratonovich or a Stratonovich-like convention. The It\^o-convention clearly does not give the Boltzmann distribution for equilibrium and it gives the MBD. The dilemma of It\^o-vs-Stratonovich conventions is apparently resolved about 40 years ago by the understanding of the fact that different conventions represent different physical processes \cite{van1981ito,mannella2022ito}. Therefore, the sole purpose in the context of thermal equilibrium should be to identify which convention corresponds to a process which is consistent with the demands of a thermal equilibrium. One may also note, in this context, that the Stratonovich or Stratonovich-like conventions involve correlated (anticipating) noise. There have also been attempts to look into this issue in some alternative way \cite{tupper2012paradox} where it is argued that specification of coordinate dependence of diffusivity needs to be accompanied by additional information to specify the state. The It\^o, Stratonovich and H\"anggi-Klimontovich conventions have been related to an infinite density and its shape by Leibovich and Barkai in \cite{leibovich2019infinite} to shed light on various interpretations.

In the present paper, we focus on the results of a classic experiment reported in 1994 by Faucheux and Libchaber \cite{faucheux1994confined} to compare the BD and the MBD with the results of this experiment. We particularly choose this experimental result for comparison of the BD with the MBD because, {\it one can do this comparison here without involving any free (adjustable) parameter for curve fitting}. When two different functions are compared with experimental outcome, ideally, one must not use any free parameter for curve fitting and this experimental result of \cite{faucheux1994confined} gives us that opportunity. The comparison shows good matching of the experimental results of average coordinate-dependent diffusivity with those computed using the MBD.

To our knowledge, it is the first experiment that had attempted to find the trajectory of a BP by direct visualization of the particle \cite{faucheux1994confined}. Based on this direct tracking, the authors have made an effort to reconcile the position-dependent-diffusivity to the Stokes-Einstein relation i.e., they have considered thermal equilibration of the BP. Although, the BD had been used, however, there are observations reported in the same paper that the BD could not fit experimental data in the way it should have fitted it. Moreover, each and every parameter of the system is experimentally determined and reported in the paper for each and every particle tracked \cite{faucheux1994confined}. 

The plan of the present paper goes in the following way. In the next section, we would first describe the experiment that has been done by Faucheux and Libchaber and we give a description of the theoretical methods used by these authors. In the next section we will describe the details of comparison of the experimental results with the BD and the MBD where the theory practically remains the same as the one used by Faucheux and Libchaber except for the method of their theoretical curve fitting. Following that, we conclude the paper with a discussion.

\section{The experiment and theoretical method of Faucheux and Libchaber}
\subsection{The experiment}
In the experiment of Faucheux and Libchaber \cite{faucheux1994confined}, they observed Brownian motion of silica and latex beads (considered spheres) of diameter 1 to 3 $\mu$m in ultra-pure water confined within two clean horizontal glass plates of spacing between 6 to 1000 $\mu$m. The real-time position of the diffusing spheres on a horizontal plane ($x$ and $y$ coordinate) were taken by a CCD camera coupled to the microscope. A time series of about 3 minute were constructed for each of these beads where the typical diffusion coefficient is about 1 $\mu$m$^2$/s. Typically, 12 such time series for each bead have been considered for an ensemble average which corresponds to about half an hour of recording for each bead. The typical time over which the water gets contaminated in this experiment was assessed to be about a day.

Based on these time series of the horizontal positions of individual beads the diffusivity $D_{||}$ of the particles were determined over a horizontal plane which actually is a projection of the 3-dimensional trajectory of the bead on a plane perpendicular to the vertical. The diffusivity on the horizontal plane $D_{||}$ and the diffusivity of the bead in the vertical ($z$-direction) $D_\perp$ are functions of the distance of the bead from the confining horizontal glass plates. Therefore, the horizontal diffusivity of a bead measured in this projection is an average over the vertical direction ($z$-coordinate) where the probability density for position distribution could be the BD or the MBD which is to be found out. Care was taken in this experiment to avoid the beads coming too close to each other and interact hydro-dynamically such that the diffusion remains only a function of the vertical $z$-coordinate. For other details of the experimental determination of the $D_{||}$ refer Faucheux and Libchaber \cite{faucheux1994confined}.

\begin{figure}
    \centering
    \includegraphics[width= .46\textwidth]{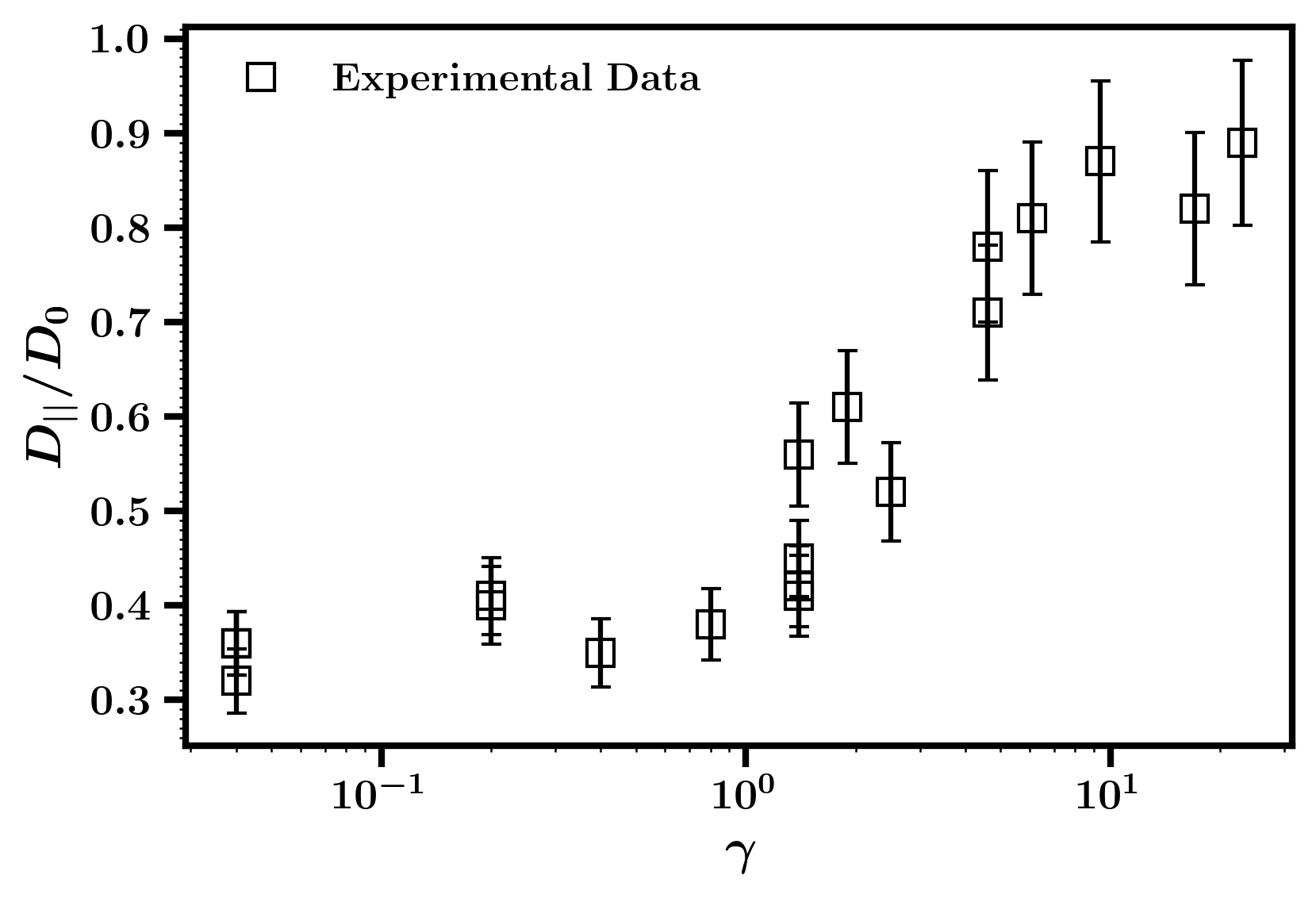}
    \caption{Experimentally determined ratio $D_{||}/D_0$ as a function of $\gamma$}
    \label{fig:my_label}
\end{figure}

\subsection{Coordinate dependence of horizontal-diffusivity on the distance from wall}
Among these beads, the heavier ones would have a smaller average distance from the lower glass plate where the lighter ones would have average height relatively larger. The diffusive excursion of these beads will be around these average height in such a way that the average height remains stationary under steady-state (thermal equilibrium) conditions. Keeping this in mind, Faucheux and Libchaber used a parameter defined as
\begin{equation}
    \gamma = \frac{h-r}{r}
\end{equation}
where $r$ is the radius of the bead and the average vertical height $h$ is defined as
\begin{eqnarray}\nonumber
h &=& \int_r^{t-r}{zP_B(z)dz}\\
&=& \frac{e^{-r/L}(rL+L^2)-e^{(r-t)/L}([t-r]L+L^2)}{L(e^{-r/L}-e^{(r-t)/L})},
\end{eqnarray}
where 
\begin{equation}
    P_B(z) = \left (\frac{1}{L}\right )\frac{e^{-z/L}}{e^{-r/L}-e^{(r-t)/L}},
\end{equation}
\begin{equation}
    L=k_BT/g\bigtriangleup\!m 
\end{equation}
and 
\begin{equation}
    \bigtriangleup m = \frac{4}{3}\pi r^3(\rho-\rho_0),
\end{equation}
to plot the experimentally obtained $D_{||}$ against this parameter. This is a nice choice of a parameter for the horizontal diffusivity $D_{||}$ to be plotted against because there is vertical excursion of the bead associated to its horizontal motion and its vertical as well as horizontal diffusivity are functions of the vertical distance of the bead from the confining plates. However, in the experimentally evaluated $D_{||}$ this dependence of $z$-coordinate appears on an average in this $z$-direction because $D_{||}$ is evaluated from a projection on the $x-y$ plane. In the above expressions $\rho$ is the density of the material of the bead, $\rho_0$ is the density of water considered in the present paper to be $1000$ kg/m$^3$ at room temperature $T=300$ K, $g$ is the acceleration due to gravity taken to be $9.8$~ m/s$^2$, $t$ is the spacing between the confining plates and $z$ is the vertical coordinate of the centre of mass of the bead taken to be a sphere. 

The experimentally measured $D_{||}$ normalized by the bulk diffusivity $D_0$ is then plotted against this parameter $\gamma$ (following \cite{faucheux1994confined}) in Fig.1. In this figure, the data-points (as given in the Table I of \cite{faucheux1994confined}) are plotted by the symbol of square boxes and the error-bars are extracted from the Fig.4 of \cite{faucheux1994confined} using WebPlotDigitizer (https://apps.automeris.io/wpd/). In what follows, the same box symbol and error-bars would be used in all the figures to correspond to the experimental data-points. Relevant data from the Table I of the paper \cite{faucheux1994confined} are shown in the Table I of the present paper.
\begin{widetext}

\begin{table*}[h!]
    \centering
    \resizebox{12cm}{!}{%
    \begin{tabular}{|c|c|c|c|c|c|c|}
    \hline  \textbf{\makecell{Sample \\ No.}} &   \textbf{\makecell{Bead \\ diameter \\ ($\mu$m)}} &  \textbf{\makecell{Bead \\ density \\ (\bm{$g/cm^3$})}} &  \textbf{\makecell{L \\ ($\mu$m)}} &  \textbf{\makecell{Sample \\ thickness t \\ ($\mu$m)}} & \textbf{\makecell{ \bm{$D_{||}/D_0$} \\ (expt)}} &
    \textbf{\makecell{\bm{$\gamma$}}} \\ \hline
        1 & 2.5 & 2.1 & 0.05 & 6 & 0.32 & 0.04\\ 
        2 & 2.5 & 2.1 & 0.05 & 25 & 0.36 & 0.04\\
        3 & 2.5 & 2.1 & 0.05 & 100 & 0.36 & 0.04\\ 
        4 & 2.5 & 2.1 & 0.05 & 1000 & 0.36 & 0.04\\ 
        5 & 3.5 & 1.05 & 0.37 & 12 & 0.4 & 0.2\\ 
        6 & 3.5 & 1.05 & 0.37 & 50 & 0.41 & 0.2\\ 
        7 & 3 & 1.05 & 0.6 & 12 & 0.35 & 0.4\\ 
        8 & 2.5 & 1.05 & 1.03 & 12 & 0.38 & 0.8\\ 
        9 & 2 & 1.05 & 2.02 & 6 & 0.45 & 1.4\\ 
        10 & 1 & 2.1 & 0.73 & 6 & 0.41 & 1.4\\ 
        11 & 1 & 2.1 & 0.73 & 25 & 0.42 & 1.4\\ 
        12 & 1 & 2.1 & 0.73 & 100 & 0.56 & 1.4 \\ 
        13 & 2 & 1.05 & 2.02 & 12 & 0.61 & 1.9 \\ 
        14 & 1.5 & 1.05 & 4.8 & 6 & 0.52 & 2.5 \\ 
        15 & 1.5 & 1.05 & 4.8 & 12 & 0.78 & 4.6 \\ 
        16 & 1 & 1.05 & 16.1 & 6 & 0.71 & 4.6 \\ 
        17 & 1.5 & 1.05 & 4.78 & 25 & 0.81 & 6.1 \\ 
        18 & 1 & 1.05 & 16.1 & 12 & 0.87 & 9.4 \\ 
        19 & 1 & 1.05 & 16.1 & 25 & 0.82 & 17 \\ 
        20 & 1 & 1.05 & 16.1 & 50 & 0.89 & 23 \\ \hline
    \end{tabular}%
    }
    \caption{This table contains the relevant data as is given in the Table 1 of \cite{faucheux1994confined}. This data is used to plot $D_{||}/D_0$ vs $\gamma$ in Fig.1.}
\end{table*}

\end{widetext}

\subsection{Effect of the wall on the damping and diffusivity}
The effect of the wall on the damping coefficients (as considered by Faucheux and Libchaber \cite{Faxen1924,Brenner1961TheSM,Cox1967TheSM,goldman1967slow}) are well known with support from molecular-dynamics \cite{chio2020hindered}. 
\begin{eqnarray}\nonumber
    &\eta_x&=\eta_y\simeq\\\nonumber
    &\eta_0&[1-\frac{9}{16}(r/z)+\frac{1}{8}(r/z)^3-\frac{45}{256}(r/z)^4-\frac{1}{16}(r/z)^5]^{-1},\\
\end{eqnarray}
and
\begin{eqnarray}\nonumber
    \eta_z &=&\frac{4}{3}\eta_0\sinh{\alpha}\sum_{n=1}^\infty \frac{n(n+1)}{(2n-1)(2n+3)}\times\\\nonumber &&\left [ \frac{2\sinh{(2n+1)}\alpha+(2n+1)\sinh{2\alpha}}{4\sinh^2{(n+1/2)\alpha}-(2n+1)^2\sinh^2{\alpha}}-1\right ],\\
\end{eqnarray}
where $\alpha=\cosh^{-1}(z/r)$ and $\eta_0$ is the damping coefficient in the bulk of the fluid which we will consider in this paper to have a value $\eta_0=8.50\times 10^{-4}$ Pa.s. The relation (7) is exact and relation (6) is accurate up to the order $(r/z)^5$. In this paper, we present results of computations up to $n=5$ while using (7) to keep parity with the order of accuracy with (6), however, we have checked that going to a few higher orders in $n$ does not change the result appreciably.

\begin{figure}[h!]
  \subfloat[]{
	   \includegraphics[width=0.46\textwidth]{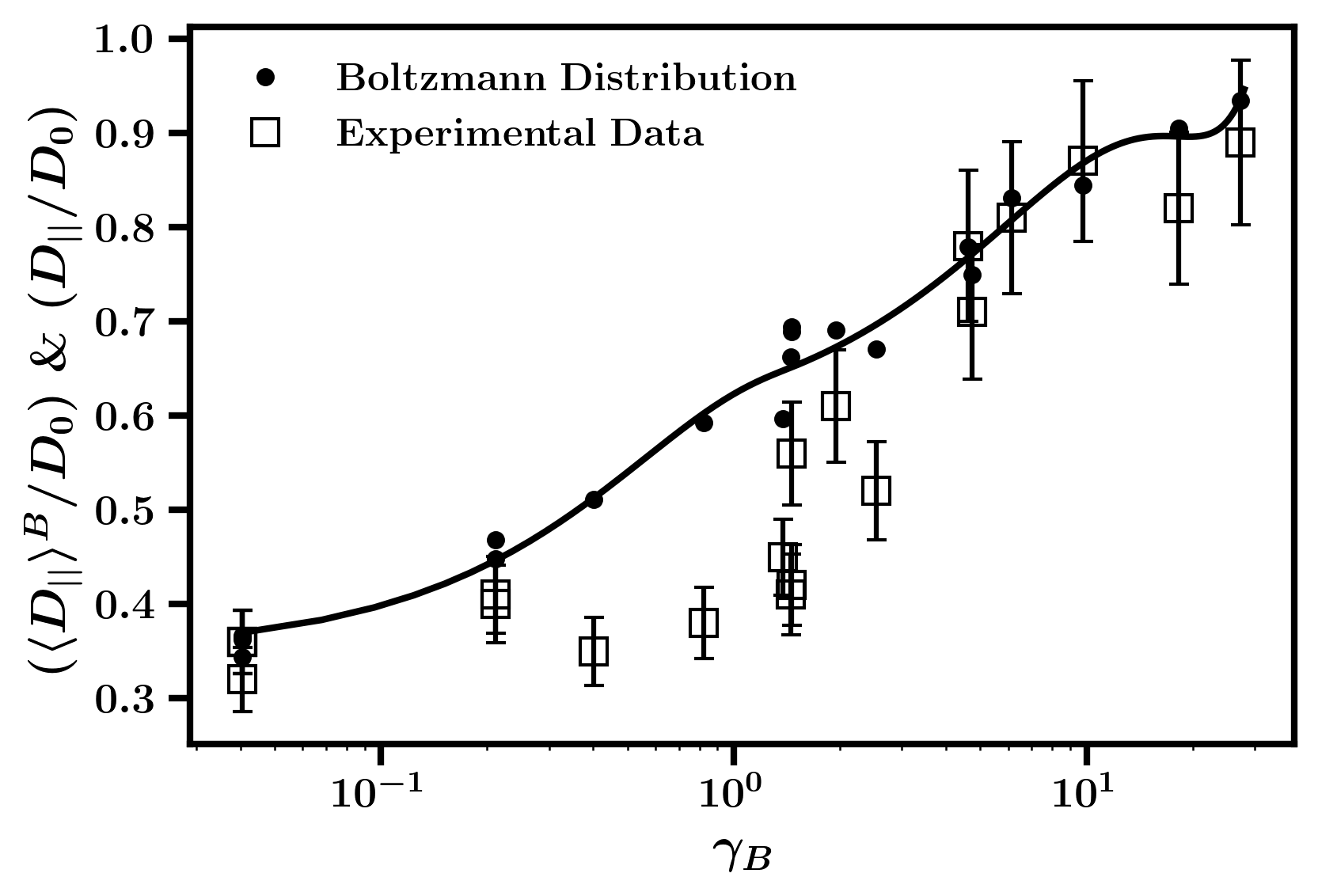}}
\\
  \subfloat[]{
	   \includegraphics[width=0.46\textwidth]{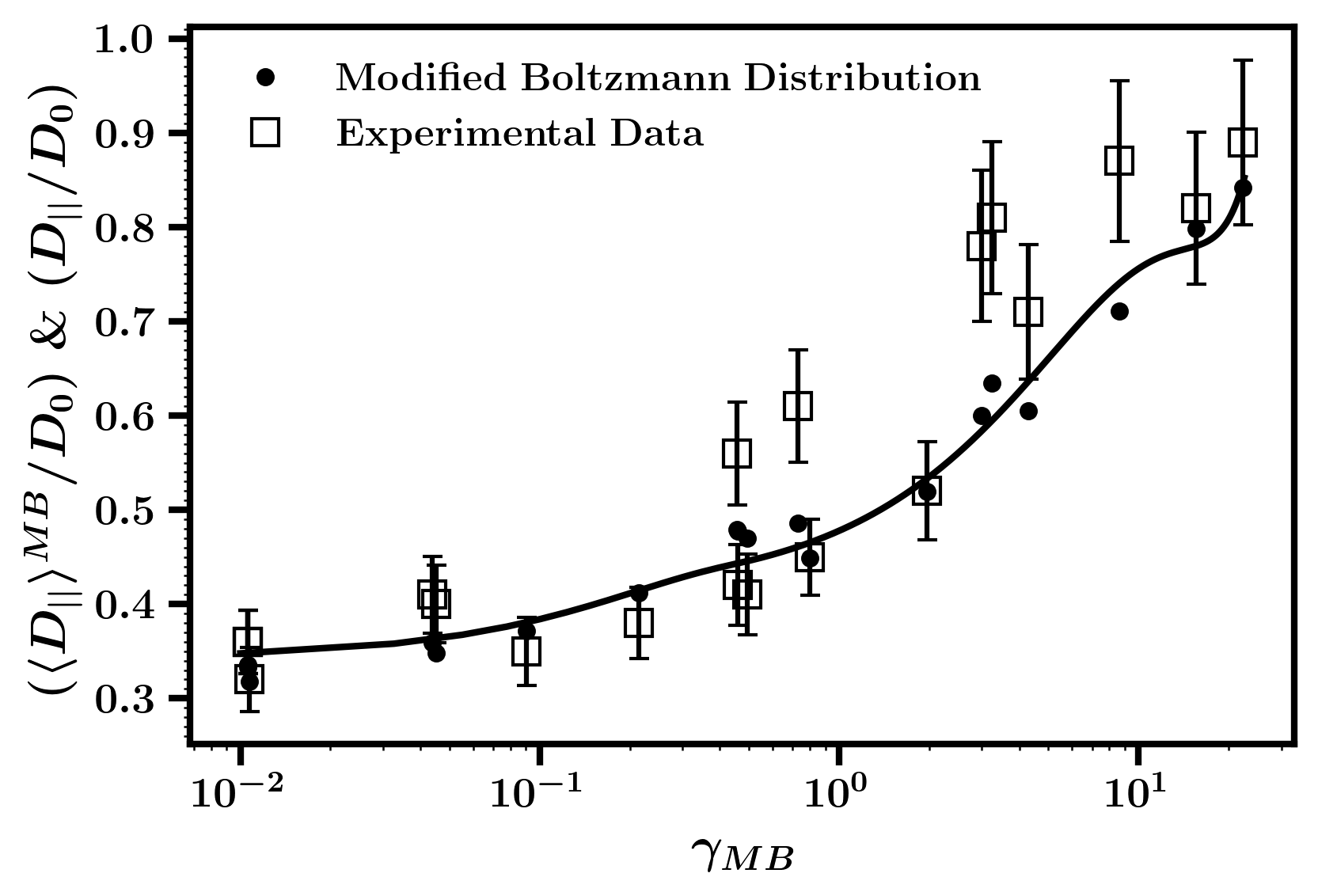}}
	
\caption{Theoretically calculated $\langle D_{||}\rangle^{B/MB}/D_0$ compared with experimental data points using $\gamma_{B/MB}$ as parameters. Theoretical data points are fitted using cubic splines with smoothing parameter as 0.01}
\end{figure}

Theoretical diffusivity on the horizontal plane is thus considered to be $D_{h}=k_BT/6\pi\eta_x r=k_BT/6\pi\eta_y r$ and that in the vertical direction is $D_\perp=k_BT/6\pi\eta_z r$, where $D_{h}=D_x=D_y$ and $D_\perp=D_z$. These expressions of the theoretical diffusivities follow from the standard fluctuation-dissipation relation or the Stokes-Einstein relation under thermal equilibrium conditions. In this particular experiment, the vertical separation ($t$) between the confining horizontal glass plates vary from 6 $\mu$m to 1000 $\mu$m. It requires to take into account the presence of both the plates and the vertical distance of the bead from them when the separation $t$ is small. Taking into account the effect of both the plates at the bottom and at the top (upper plate), Faucheux and Libchaber \cite{faucheux1994confined} introduced an effective $\eta_{x,y,z}$ as
\begin{equation}
    \eta_{x,y,z}=\eta^b_{x,y,z}+\eta^u_{x,y,z}-\eta_0,
\end{equation}
where the superscripts $b$ and $u$ indicate damping coefficients due to bottom and upper (top) plates respectively. Note that, $\eta^b_{x,y,z}$ is a function of $z/r$ where $\eta^u_{x,y,z}$ is similarly a function of $(t-z)/r$. This expression of damping coefficient $\eta_{x,y,z}$ has nice limits. When the upper plate goes to infinity $(t\to \infty)$, we get $\eta^u_{x,y,z}=\eta_0$ and only the lower plate has any effect on the bead. Thus, even when the bottom and the top plates are far away from the BP we get $\eta_{x,y,z}=\eta_0$. 

The relation (8) takes into account the independent presence of the bottom and the top plate on the damping coefficient of the BP without considering any interaction. Imagine that, only the bottom plate is present and the damping coefficient is modified due only to its presence such that
\begin{equation}\nonumber
    \eta_{x,y,z}=\eta_0 +\bigtriangleup\!\eta^b(z),
\end{equation}
where $\bigtriangleup\!\eta^b(z)$ is the modification of the damping due to the presence of the bottom plate on top of what it is there in the bulk. Similarly, when one takes into account the presence of the top plate whose hydrodynamic effect is not going to interfere with that produced by the bottom plate, there would also be an increase in the damping by the amount $\bigtriangleup\!\eta^u(z)$ on top of what the bottom plate has produced. Therefore the new damping coefficient considering the presence of both the plates is
\begin{eqnarray}\nonumber
    \eta_{x,y,z} &=& \eta_0 +\bigtriangleup\!\eta^b(z) + \bigtriangleup\!\eta^u(z)\\\nonumber
    &=& [\eta_0 +\bigtriangleup\!\eta^b(z)] + [\eta_0+ \bigtriangleup\!\eta^u(z)] -\eta_0\\\nonumber
    &=& \eta^b_{x,y,z}+\eta^u_{x,y,z}-\eta_0
\end{eqnarray}
We have found this expression (8) to work quite well and in what follows we would go by the relation (8) to take into account the hydrodynamic effects due to both the walls.


\section{Comparison of BD and MBD with the experimental result}
In this section we compare the analytically obtained height-averaged horizontal diffusivity $\langle D_{||}\rangle^{B/MB}$ with the experimentally obtained results ($D_{||}$) of Faucheux and Libchaber \cite{faucheux1994confined}. We use two different distributions the BD and the MBD to get the height-average ($z$-averaged) horizontal diffusivity in the following way
\begin{eqnarray}
    \langle D_{||}\rangle^{B}&=&\int_{r+\epsilon}^{t-(r+\epsilon)}{D_{h}(z)P_{B}(z)dz}\\
     \langle D_{||}\rangle^{MB}&=&\int_{r+\epsilon}^{t-(r+\epsilon)}{D_{h}(z)P_{MB}(z)dz},
\end{eqnarray}

The distributions are
\begin{eqnarray}
    P_{B}&=&N_{B}\exp{(-z/L)}\\
    P_{MB}&=&\frac{N_{MB}}{D_\perp(z)}\exp{(-z/L)},
\end{eqnarray}
where $N_B$ and $N_{MB}$ are the normalization constants for the BD and the MBD respectively, evaluated exactly over the same lower/upper limits for a particular bead as shown in (9) and (10). We implement a cut-off distance $\epsilon=10^{-3} \mu$m to avoid the divergence appearing in the MBD due to the vanishing of $D_\perp(z)$ at $z=r$ and $z=t-r$. We implement the same cut-off in (9) and (10) in order to make the averages with respect to the BD and the MBD identically computed. The integration here is a left-Riemann-sum with bin-width $10^{-3} \mu$m which is being presented. We have checked that the left and right Riemann-sum are giving excellent match with this bin-size and this is to be checked because of the divergence near the walls.
\begin{widetext}
\begin{table*}[!ht]
    \centering
    \resizebox{12cm}{!}{%
    \begin{tabular}{|c|c|c|c|c|c|c|c|c|c|}
    \hline  \textbf{\makecell{Sample \\ No.}} &   \textbf{\makecell{\bm{$h_B$} \\ ($\mu$m)}} & 
    \textbf{\makecell{\bm{$h_{MB}$} \\ ($\mu$m)}} & 
    \textbf{\makecell{\bm{$\gamma_{B}$}}} & 
    \textbf{\makecell{\bm{$\gamma_{MB}$}}} & 
    \textbf{\makecell{\bm{$\Gamma_{MB}$}}} & 
    \textbf{\makecell{$tL/100$ \\ \bm{$\mu$ $m^2$}}} &
    \textbf{\makecell{\bm{$\langle D_{||}\rangle^{B}/D_0$}}} &
    \textbf{\makecell{\bm{$\langle D_{||}\rangle^{MB}/D_0$}}} &
    \textbf{\makecell{\bm{$D_{||}/D_0$} \\ (expt)}} \\ \hline
        1 & 1.30 & 1.26 & 0.04 & 0.01 & 0.01 & 0.003 & 0.34 & 0.32 & 0.32 \\ 
        2 & 1.30 & 1.26 & 0.04 & 0.01 & 0.01 & 0.0125 & 0.36 & 0.33 & 0.36 \\ 
        3 & 1.30 & 1.26 & 0.04 & 0.01 & 0.01 & 0.05 & 0.37 & 0.34 & 0.36 \\ 
        4 & 1.30 & 1.26 & 0.04 & 0.01 & 0.01 & 0.5 & 0.37 & 0.34 & 0.36 \\ 
        5 & 2.12 & 1.83 & 0.21 & 0.04 & 0.04 & 0.0444 & 0.45 & 0.35 & 0.4 \\ 
        6 & 2.12 & 1.83 & 0.21 & 0.04 & 0.04 & 0.185 & 0.47 & 0.36 & 0.41 \\ 
        7 & 2.10 & 1.64 & 0.40 & 0.09 & 0.09 & 0.072 & 0.51 & 0.37 & 0.35 \\ 
        8 & 2.28 & 1.52 & 0.82 & 0.21 & 0.21 & 0.1236 & 0.59 & 0.41 & 0.38 \\ 
        9 & 2.38 & 1.80 & 1.38 & 0.80 & 0.69 & 0.1212 & 0.60 & 0.45 & 0.45 \\ 
        10 & 1.23 & 0.75 & 1.45 & 0.49 & 0.47 & 0.0438 & 0.66 & 0.47 & 0.41 \\ 
        11 & 1.23 & 0.73 & 1.46 & 0.46 & 0.45 & 0.1825 & 0.69 & 0.48 & 0.42 \\ 
        12 & 1.23 & 0.73 & 1.46 & 0.46 & 0.45 & 0.73 & 0.69 & 0.48 & 0.56 \\ 
        13 & 2.95 & 1.73 & 1.95 & 0.73 & 0.68 & 0.2424 & 0.69 & 0.49 & 0.61 \\ 
        14 & 2.65 & 2.23 & 2.54 & 1.97 & 1.48 & 0.288 & 0.67 & 0.52 & 0.52 \\ 
        15 & 4.22 & 2.99 & 4.63 & 2.99 & 2.43 & 0.576 & 0.78 & 0.60 & 0.78 \\ 
        16 & 2.87 & 2.64 & 4.74 & 4.28 & 2.75 & 0.966 & 0.75 & 0.60 & 0.71 \\ 
        17 & 5.36 & 3.18 & 6.14 & 3.24 & 2.93 & 1.195 & 0.83 & 0.63 & 0.81 \\ 
        18 & 5.38 & 4.82 & 9.76 & 8.63 & 5.53 & 1.932 & 0.84 & 0.71 & 0.87 \\ 
        19 & 9.62 & 8.28 & 18.25 & 15.56 & 10.72 & 4.025 & 0.90 & 0.80 & 0.82 \\ 
        20 & 14.15 & 11.64 & 27.30 & 22.29 & 17.32 & 8.05 & 0.93 & 0.84 & 0.89 \\ \hline
    \end{tabular}%
    }
    \caption{Contains values of different parameters ($\gamma_B$, $\gamma_{MB}$, $\Gamma_{MB}$ and $tL/100$) and diffusivity ratios, both theoretical ($\langle D_{||}\rangle^{B/MB}/D_0$) and experimental $D_{||}/D_0$. Also, $h_B$ and $h_{MB}$ are provided as they are used to calculate the parameters.}
\end{table*}
\end{widetext}
It is important to note that, for a matching of the experimental results with the theoretical ones (9) and (10), we need to find out the theoretical values of the average diffusion coefficients for each individual particle. The corresponding parameter $\gamma$ has also to be found out for each individual particle because the boundary conditions, matter density and size are quite different for different particle which actually correspond to different experimental conditions. In Fig.2(a) and (b), we have compared the theoretical values (according to (9) and (10)) of the average diffusivity of individual particles against the experimental data using the parameter $\gamma_B$ and $\gamma_{MB}$. These are the same $\gamma$ parameter of Faucheux and Libchaber, but, computed by us within the specified limits using BD and MBD respectively. Theoretical data points for Fig.2 and Fig.3 are given in Table II. The continuous line in each figure corresponds to the cubic-spline to the theoretical data points with a smoothing factor $0.01$. As we can see that, the parameters $\gamma_{B/MB}$ are distribution dependent, and that is why we have plotted two different graphs and have not overlaid the theoretical diffusivities on the same graph.

Faucheux and Libchaber have pointed out in ref.\cite{faucheux1994confined} that the BD does not work to match the experimental data when used as in (9). Fig.2(a) shows that this assessment of Faucheux and Libchaber is correct and the Boltzmann distribution fails to match the trend of the experimental result particularly in the region of lower values of $D_{||}$ where the modification of the diffusivity is appreciable due to the proximity of the wall. The theoretical and experimental values are quite close while the BD is used at the two ends i.e., for very heavy particles which practically stay at the bottom and the ones which are very light and are mostly in the bulk. Note that, very heavy beads which stay near the bottom plate they do not in general undergo much vertical excursion and as a result the effect of averaging in the vertical direction is quite small for these particles. Moreover, very light ones are staying mostly in the bulk of the fluid where the diffusivity is less perturbed and thereby the effect of average in the vertical direction would have little effect on those as well. It is the trend in the intermediate region which is the most important part. However, in this middle region, the trend of the variation of the experimental result is missed by $\langle D_{||}\rangle^B$.

On the contrary, Fig.2(b) shows a much better match between the theoretical and experimental result in the intermediate region when the MBD is used. It also appears that the MBD is not matching the larger diffusivity region as much as the BD does. Note that, the error bars in the larger diffusivity region are larger due to limitations in the averaging process over the experimental trajectories mentioned in \cite{faucheux1994confined}. Let us look at the data of the larger diffusivity (near bulk) more closely. For example, the sample number 18 in Table I, which is the third experimental data-point in the plots from the right hand side end is clearly an over-count. This particle has a bead diameter 1 $\mu$m which is the same as the bead of the sample number 19 and both have the same density. Where the sample number 18 has a sample thickness $t=12$ $\mu$m that of the sample number 19 is 25. Given this, the $D_{||}/D_0$ of sample number 18 should be smaller than that of the sample number 19, however, that is not the experimental result as shown in Table I. Similarly, if one compares the experimental value of $D_{||}/D_0$ of sample number 17 with that of the sample number 19, one would find that the experimental value is larger (quite close to that of sample number 19) than that it should be for the sample number 17 given the bigger diameter of the bead. The experimental data in Table I, in this way, reveals that there apparently exists some bias towards over-counting the diffusivity of the lighter particles.

\begin{figure}[h!]
  \subfloat[]{
	   \includegraphics[width=0.46\textwidth]{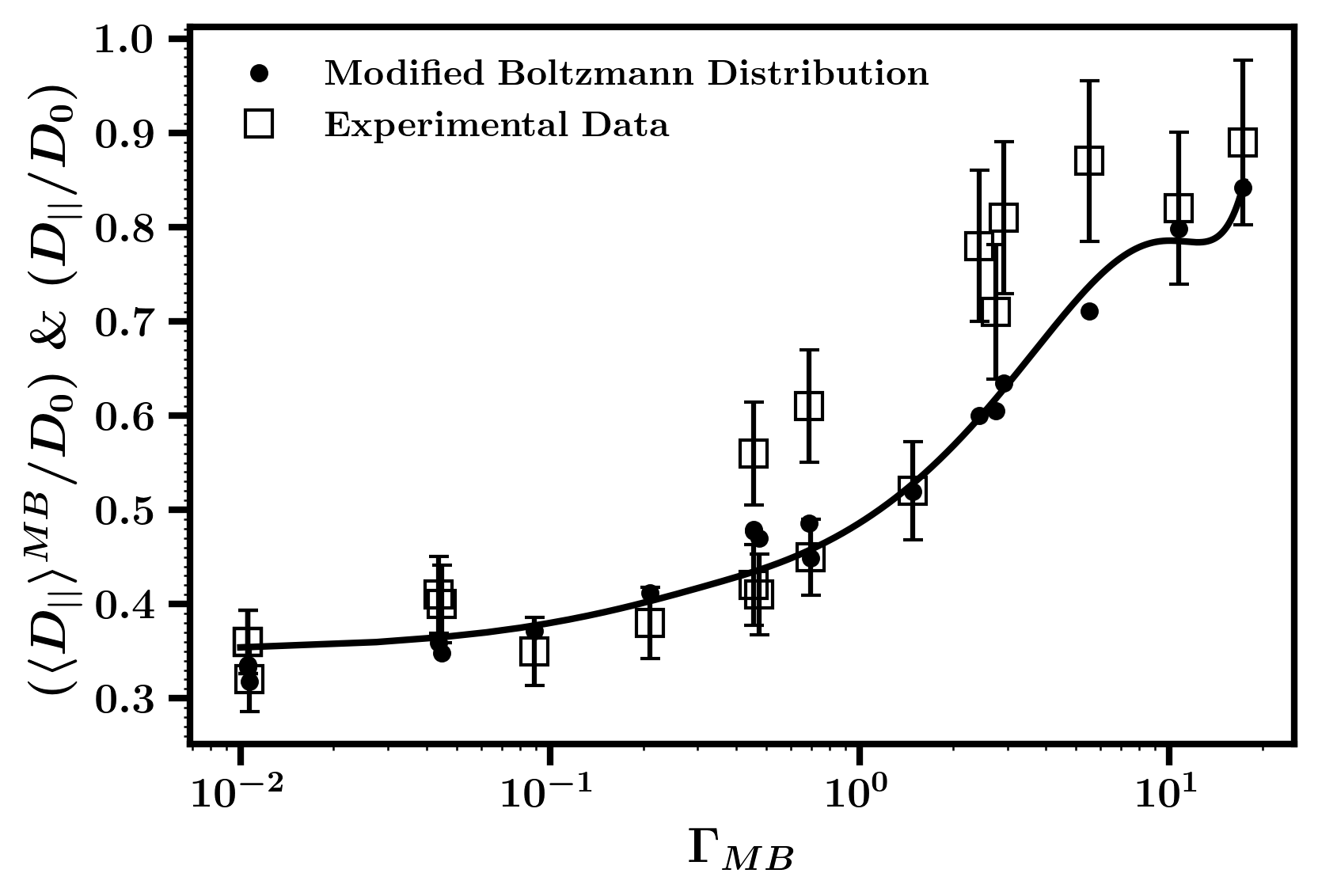}}
\\
  \subfloat[]{
	   \includegraphics[width=0.46\textwidth]{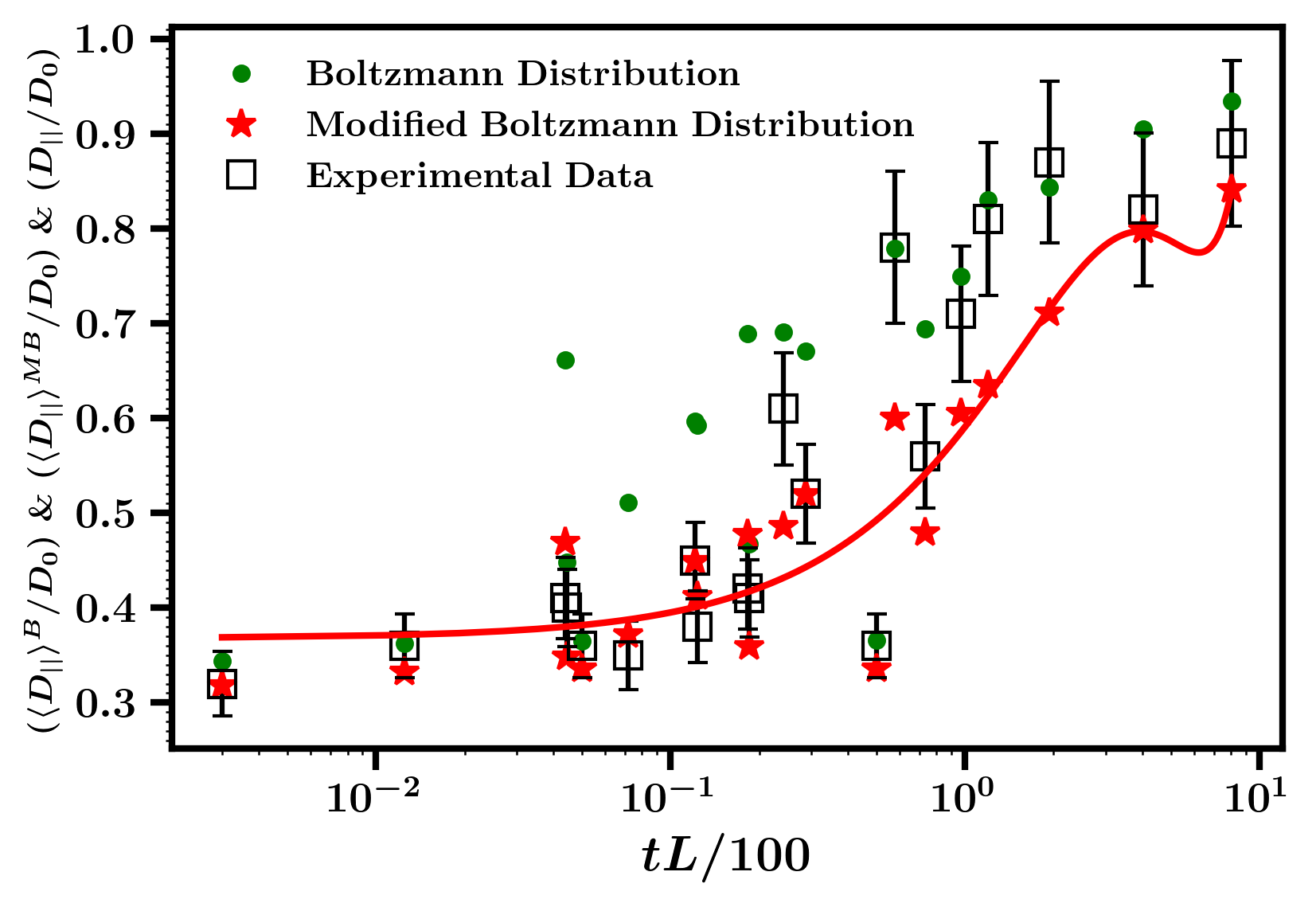}}
	
\caption{Theoretical $\langle D_{||}\rangle^{MB}/D_0$ compared with experimental data points and $\langle D_{||}\rangle^{B}/D_0$ using parameters $\Gamma_{MB}$ and $tL/100$, respectively, in (a) the cubic spline to the theoretical result has a smoothing factor 0.01 and in (b) A cubic spline with smoothing factor 0.1 is drawn for $\langle D_{||}\rangle^{MB}/D_0$.}
\end{figure}

	

In the Fig.3(a) and (b), we compare the experimental results of Faucheux and Libchaber \cite{faucheux1994confined} with the $\langle D_{||}\rangle^{MB}$ where we use two different parameters $\Gamma_{MB}$ and $tL/100$, respectively. The latter parameter is a distribution independent parameter and we can overlay graphs corresponding to the BD and the MBD here. We have used the modification of the Faucheux and Libchaber's parameter as $\Gamma_{MB}$ (in Fig.3(a)) to take into account the distance of the particle from both the lower and the upper confining planes in the parameter as
\begin{equation}
    \Gamma_{MB}=\gamma_{MB}\gamma^\prime_{MB},
\end{equation}
where
\begin{equation}
    \gamma^\prime_{MB}=1-\gamma_{MB}\frac{r}{t},
\end{equation}
where $h$ is evaluated by averaging with MBD. $\gamma^\prime_{MB}$ is a fraction which has the limit of unity as the upper plate goes far away $(t\to\infty)$ and the effect of the lower plate only is there on the parameter at that limit. Otherwise, the presence of the factor $\gamma^\prime_{MB}$ lowers the value of the parameter $\Gamma_{MB}$ in comparison with $\gamma_{MB}$ depending upon the proximity of the upper plate. As has already been pointed out by Faucheux and Libchaber, effect of the upper plate in the parameter $\gamma$ is not felt that much is clear from the comparison of the Fig.2(b) and Fig.3(a). 

The parameter in Fig.3(b) is $tL/100$ (normalization is done to have similar scales in all the figures) which is distribution independent. In this figure we have plotted $\langle D_{||}\rangle^B$ by green dots and the $\langle D_{||}\rangle^{MB}$ by red stars. We have drawn the cubic spline to only the $\langle D_{||}\rangle^{MB}$ and not for $\langle D_{||}\rangle^{B}$ because the data points are too scattered. This figure clearly shows us that the matching of the theory with experimental data remains quite good with the MBD for the beads staying closer to the lower plate (and hence feeling more hydrodynamic effects) in comparison to those which are lighter. However, as we have already discussed above, from the table of experimental data points for $D_{||}$ as is given by Faucheux and Libchaber, there probably exists a trend of a little over-counting for the $D_{||}$ of the lighter particles and this could be resolved by a similar experiment if done at present times. 

Corresponding to the parameter $tL/100$ the sample number 2, 3 and 4 sit at the parameter values $tl/100 = 0.0125$, $0.05$ and $0.5$. However, all these samples being heaviest particles have the same experimentally measured $D_{||}/D_0$ indicating that the sample thickness $t$ beyond $t=25$ $\mu$m does not have any effect on the diffusivity of these particles because those are anyway very close to the bottom wall. However, the parameter $tL/100$ is making these sample points sit at different points on the horizontal axis because the parameter counts the sample thickness even where it is not influencing the result. Apart from this, the parameter $tL/100$ appears to be a good parameter for almost all other sample numbers to directly compare the effects of the BD and the MBD with the experimental results.

\section{Discussion}
The thermal equilibrium distribution of a BP with coordinate dependent diffusion is the MBD has been theoretically argued for in the references \cite{bhattacharyay2019equilibrium,bhattacharyay2020generalization}. In the reference \cite{maniar2021random} it has been shown that this coordinate-dependent diffusion is indeed a continuum limit of a random walk in contact with a heat-bath. The basis of the theoretical argument in favour of an It\^o-process as representing thermal equilibrium has been identifying the It\^o-process to be consistent with the local applicability of the Stokes-Einstein relation (fluctuation-dissipation relation). Moreover, an It\^o-process does not introduce any correlation in the thermal noise which is unavoidable when the Stratonovich or a Stratonovich-like conventions are used. 

A comparison of the It\^o-process with existing experimental result can really set the stage for further serious exploration in this direction. There are number of ways in which the BD is theoretically imposed for equilibrium using the Stratonovich or a Stratonovich-like convention in such a multiplicative noise problems. These theories are inconsistent (despite imposing the BD) from the stand point of fundamental requirements of a stochastic process to be a thermal equilibrium. We have given here an alternative explanation to the experimental results of \cite{faucheux1994confined} which had an explanation already given by Faucheux and Libchaber using the BD. Moreover, it is apparent from this comparison that the MBD (corresponding to an It\^o-process) better captures the experimental results than the BD.

It is crucial for fundamental reasons that the equilibrium distribution of a BP with coordinate-dependent (or state-dependent for a many-body system) diffusion be experimentally settled. Coordinate- or state-dependent diffusion has the potential to generalize our knowledge of thermal equilibrium to a great extent because this process takes into account an in-homogeneity of space keeping intact the thermal homogeneity of the heat-bath. In the conventional theory of uniform diffusion, this in-homogeneity of space is absent. In the presence of coordinate-dependent diffusion the homogeneity of space (even in the absence of a conservative force) gets broken and decouples from the thermal homogeneity of the heat-bath so that the particle can see a constant temperature in an otherwise in-homogeneous space. This decoupling can result in counter-intuitive reality to our present understanding of thermal equilibrium.

Diffusion is a major source of transport in numerous biological systems and the space remains quite crowded in such cases where the coordinate dependence of the diffusion could be the rule and not an exception. Moreover, coordinate dependence of diffusion could have other counter-intuitive effects of producing directed motions in structured objects as is shown in \cite{bhattacharyay2012directed,sharma2020conversion}. These two effects of coordinate dependence of diffusion i.e. modifying the Boltzmann distribution and spontaneously filtering of Brownian motion by structured objects in the presence of state dependent diffusion may prove to be essential ingredients for our understanding of a plethora of other phenomena that happen at meso-scales being driven by equilibrium fluctuations.

The experimental results of \cite{faucheux1994confined} is very reliable because it has used ultra-pure water as the fluid. The use of ultra-pure water and other experimental cares taken to keep the fluid pure (or doing the experiment within the time the fluid remains contaminant-free) has kept the potential term for the Boltzmann factor to be in its simplest form. Practically no amount of free ions were presumably present in the fluid helping avoid the the presence of any Debye-Huckel term. This is the reason that there is no need to deal with any {\it free parameter} in this context. The experiment was done about 30 years ago, however, still it remains ever more important because of the way it was done. At present, the particle tracking ability has improved to an extremely high level of accuracy. It is a matter of time, to our belief, that the MBD could be shown to be the equilibrium distribution for coordinate-dependent diffusion as opposed to the BD once one does a similar experiment where there would be no need to use adjustable parameters to fit the experimental results.

\bibliography{reference.bib}

\end{document}